\documentclass[12pt]{article}
\usepackage{amscd,amsmath,amssymb,amstext,amsthm,exscale,latexsym}
\usepackage{graphicx}
\newcommand {\al}   {\alpha}       \newcommand {\bt}  {\beta}
\newcommand {\g }   {\gamma}       
\newcommand {\dl}   {\delta}       \newcommand {\e }  {\epsilon}
        
\newcommand {\ve}   {\varepsilon}  
\newcommand {\lm}   {\lambda}      \newcommand {\m }  {\mu}
          
\newcommand {\s }   {\sigma}       
         
\newcommand {\vf }  {\varphi}      
         \newcommand {\om}  {\omega}
      \newcommand {\Om}  {\Omega}
\newcommand {\Th}   {\Theta}       
\newcommand {\pl}   {\partial}     \newcommand {\nb}  {\nabla}
\renewcommand {\sin}{{\sf\,sin\,}}       \renewcommand {\cos}{{\sf\,cos\,}}

\newcommand {\sh}{{\sf\,sh\,}}         
\renewcommand {\tanh}{{\sf\,th\,}}

       \renewcommand {\exp}{{\sf\,exp\,}}

       \renewcommand {\lim}{{\sf\,lim\,}}
\newcommand   {\ex}{{\sf\,e}}

\newcommand   {\const}{{\sf\,const}}     
         \newcommand   {\tr}{{\sf\,tr\,}}

             \renewcommand   {\P}{{\sf P}}

\newcommand {\MC}  {{\mathbb C}}

\newcommand {\MM}  {{\mathbb M}}   
\newcommand {\MO}  {{\mathbb O}}   
   \newcommand {\MR}  {{\mathbb R}}
\newcommand {\MS}  {{\mathbb S}}   
\newcommand {\MU}  {{\mathbb U}}   
   
   \newcommand {\MZ}  {{\mathbb Z}}

\newcommand {\Go}  {\mathfrak{o}}   
   
\newcommand {\Gs}  {\mathfrak{s}}   
\newcommand {\Gu}  {\mathfrak{u}}

   \newcommand {\Bb}  {\boldsymbol{b}}

   \newcommand {\Bx}  {\boldsymbol{x}}
   
\newcommand {\CE }  {{\cal E}}

\newcommand {\Sc}  {{\textsc{c}}}

\newcommand {\Ss}  {{\textsc{s}}}

\newcommand {\one}  {1\!\!1}

\newtheorem{theorem}{Theorem}[section]

\theoremstyle{definition}

\newtheorem{exa}{Example}[section]

\begin{document}
\title     {Disclinations in the geometric theory of defects}
\author    {M. O. Katanaev
            \thanks{E-mail: katanaev@mi-ras.ru}\\ \\
            \sl Steklov Mathematical institute,\\
            \sl 119991, Moscow, Gubkina St., 8}
\date{21 March 2020}
\maketitle
\begin{abstract}
In the geometric theory of defects, media with a spin structure, for example,
ferromagnet, is considered as a manifold with given Riemann--Cartan geometry.
We consider the case with the Euclidean metric corresponding to the absence of
elastic deformations but with nontrivial $\MS\MO(3)$-connection which produces
nontrivial curvature and torsion tensors. We show that the 't Hooft--Polyakov
monopole has physical interpretation in solid state physics describing media
with continuous distribution of dislocations and disclinations. The
Chern--Simons action is used for the description of single disclinations. Two
examples of point disclinations are considered: spherically symmetric point
``hedgehog'' disclination and the point disclination for which the $n$-field
has a fixed value at infinity and essential singularity at the origin. The
example of linear disclinations with the Franc vector divisible by $2\pi$ is
considered.
\end{abstract}
\section{Introduction}
Many physical properties of solids such as plasticity, melting, growth and
others, are defined by defects in the crystallin structure. Therefore the
study of defects is the actual scientific problem important for applications.
In spite of the existence of dozens of monographs and thousands of papers, the
fundamental theory of defects is now absent.

One of the promising approach to the construction of the theory of defects is
based on Riemann--Cartan geometry which is given by nontrivial metric and
torsion. In this approach, a crystal is considered as an elastic continuous
media with a spin structure. If the displacement vector field is a smooth
function, then the crystal possesses only elastic stresses corresponding to
diffeomorphisms of the flat Euclidean space. If the displacement vector field
has discontinuities, then we say that the media has defects in the elastic
structure which are called dislocations and resulting in nontrivial geometry.
Namely, they lead to nonzero torsion tensor which is equal to the surface
density of the Burgers vector.

The idea to relate torsion to dislocations aroused if fifties [1--4].
\nocite{Kondo52,Nye53,BiBuSm55,Kroner58}
This approach is successively developed until now (we note reviews [5--11])
\nocite{SedBer67,Kleman80A,Kroner81,DzyVol88,KadEde83ER,KunKun86,Kleine89}
and often called the gauge theory of dislocations. Similar approach is also
developed in gravity \cite{HeMcMiNe95}. It is interesting, that E.~Cartan
introduced the notion of torsion in geometry using the analogy with
mechanics of elastic media \cite{Cartan22R}.

Parallel to the study of dislocations, another types of defects were intensively
investigated. The point is that many solids have not only elastic properties but
possess also a spin structure, for example, ferromagnets, liquid crystals, spin
glasses, etc. In this case, there are defects in the spin structure called
disclinations \cite{Frank58}. They arise when the $n$-field describing a spin
structure has discontinuities. The presence of disclinations is also connected to
nontrivial geometry. Namely, the curvature tensor for $\MS\MO(3)$-connection
is equal to the surface density of the Frank vector. The gauge approach based on
the rotational group $\MS\MO(3)$ was used in this case \cite{DzyVol78}.
The $\MS\MO(3)$-gauge models of spin glasses with defects were considered in
\cite{Hertz78,RivDuf82}.

The geometric theory of static distribution of defects describing both types of
defects from the unique point of view was proposed in \cite{KatVol92}. In
contrast to other approaches, the only independent variables in this case are
the vielbein and $\MS\MO(3)$-connection. Torsion and curvature tensors have
straightforward physical interpretation as surface densities of dislocations
and disclinations, respectively. Covariant equations of equilibrium are
postulated for the vielbein and $\MS\MO(3)$-connection as in gravity models with
torsion. Since any solution of equilibrium equations is defined up to general
coordinate transformations and local $\MS\MO(3)$-rotations, we have to choose
the coordinate system (to fix the gauge) to specify the unique solution. The
elastic gauge for the vielbein \cite{Katana03R} and the Lorentz gauge for the
$\MS\MO(3)$-connection \cite{Katana04R} were proposed recently. We stress that
the displacement and rotational vector fields are not considered as independent
variables in our approach. These notions can be introduced only in
those domains of media where defects are absent. In this case equilibrium
equations for vielbein and $\MS\MO(3)$-connection are identically satisfied,
the elastic gauge reduces to equations of nonlinear elasticity theory for the
displacement vector field, and the Lorentz gauge transforms into equations of
the principal chiral $\MS\MO(3)$-field. In other words, one can choose two
fundamental models: the elasticity theory and the principal chiral field, to
fix the coordinate system.

Presence of defects produces nontrivial Riemann--Cartan geometry. It means that
we should modify the equations of phenomena related directly to elastic media.
For example, if propagation of phonons in ideal crystal is described by the wave
equation, then taking into account the influence of dislocations is simple. To
this end we have to replace the Euclidean metric by the nontrivial metric
describing the distribution of defects. Scattering of phonons on straight
parallel dislocations were analysed in \cite{Moraes96,dePaMo98,KatVol99}.
For description of quantum phenomena, the same substitution of metric must be
done in the Schrodinger equation. It is shown now that the presence of
dislocations influences essentially physical phenomena (see., e.g., [24--38]).
\nocite{Azeved03,deBKat07,FuMode08,deBKat09,deKaKoSh10,Katana10,LazHeh10,%
RanHug11,BoeObu12,KatMan12AR,KatMan12B,BakFur13A,Katana15C,%
Katana16A,CiIoPaZa20}

Dislocations and their influence on physical properties of various media were
mainly considered as applications of geometric theory of defects until now. It
was assumed that the $\MS\MO(3)$-connection is trivial but the metric differs
from the Euclidean one and corresponds to a given dislocation. Disclinations in
the framework of geometric theory of defects started to be considered a short
time ago. Problems of this type imply that the metric is Euclidean (elastic
deformations are absent), but the $\MS\MO(3)$-connection is nontrivial. As far
as we know, the first such papers describe straight disclination
\cite{Katana18AR,Katana18BR}. The Chern--Simons action for the
$\MS\MO(3)$-connection was used there.

In the geometric theory of defects, the $\MS\MO(3)$-connection is used instead
of the $n$-field. To this end we need the transformation of the $n$-field to the
angular rotation field. This transformation is nontrivial because the additional
gauge degree of freedom appears \cite{Katana19R}. In addition, the
$\MS\MO(2)$-gauge model without the $\MS\MO(2)$-gauge field appears.

In this review, we consider the case of Euclidean metric but nontrivial
$\MS\MO(3)$-connection which corresponds to the presence of disclinations. The
short introduction into the geometric theory of defects is given at the
beginning.

Since the Lie algebras $\Gs\Go(3)$ and $\Gs\Gu(2)$ are isomorphic, then the
static solutions of $\MS\MU(2)$-gauge models can be considered as describing
some distribution of disclinations and, possibly, dislocations. In particular,
the 't Hooft--Polyakov monopole has straightforward physical interpretation in
the geometric theory of defects describing media with continuous distribution of
disclinations and dislocations \cite{Katana20A}. This is considered in Section
\ref{smotrp}.

The first examples of point disclinations in the geometric theory of defects are
based on the Chern--Simons action. The most general form of the trivial
spherically symmetric $\MS\MO(3)$-connection containing one arbitrary function
on radius is found for this case. In section \ref{sbnsgt}, we construct two
examples of point disclinations for different boundary conditions. The first one
describes the hedgehog disclination, and the second corresponds to the point
disclination with $n$-field taking a fixed value at infinity and having
essential singularity at the origin \cite{KatVol20}. In Section \ref{sbcvfg},
we describe straight linear disclinations in the framework of the geometric
theory of defects.
\section{Elastic deformations                                    \label{seldef}}
The elasticity theory (see., e.g.,  \cite{LanLif87R,Nowack70ER}) is the
classical part of mathematical physics which formulated during decades its
own language different from that of modern differential geometry in many
respects. In this section, we give necessary notions of elasticity theory from
the point of view of differential geometry (see., e.g.,
\cite{DuNoFo98R,KobNom6369R}).

In the equilibrium state, a body occupies some bounded domain in the Euclidean
space of observer $\MR^3$. The equilibrium state is not defined uniquely: we can
rotate or move a body as a whole. Any deformation and elastic stresses are
absent in this case since the Euclidean metric is invariant with respect to
these transformations. We denote Cartesian coordinates of a point of a body by
Latin letters $y^i$, $i=1,2,3$. After a deformation or motion, every point of a
body takes new position: $y\mapsto x$. This deformation corresponds to some
diffeomorphism of domains in the Euclidean space. In addition, the body acquires
the induced metric
\begin{equation}                                                  \label{ehfgty}
  \dl_{ij}\mapsto
  g_{ij}=\frac{\pl y^k}{\pl x^i}\frac{\pl y^l}{\pl x^j}\dl_{kl}.
\end{equation}
The difference
\begin{equation}                                                  \label{qvrfre}
  \e_{ij}(y):=\frac12\big(\dl_{ij}-g_{ij}(y)\big).
\end{equation}
is called {\em the deformation tensor} in the Cartesian coordinates. This
difference is correctly defined because tensor components are subtracted
pointwise. The definition implies that the deformation is identically zero in
the equilibrium state. It is also zero after translations and rotations of a
body as a whole.

An observer can add or subtract point coordinates before and after a deformation
because he works in the Euclidean space which has the natural affine structure.
After a deformation, a given point has new coordinates
\begin{equation}                                                  \label{eeldef}
  y^i\mapsto x^i:=y^i+u^i,
\end{equation}
where $u^i(x)$ is {\em the displacement vector field}, in the same coordinate
system.

The following terminology is used in the elasticity theory. If components of a
displacement vector field are considered as functions on initial point
coordinates $y$, then this system is called {\em Lagrangian coordinates}.
If coordinates after deformation $x$ are chosen as the independent ones, then we
say that {\em Eulerian} coordinates are chosen. The Lagrangian and Eulerian
coordinates are equivalent, if the domains of definition of point coordinates
of a body $x^i$ and $y^i$ are diffeomorphic. However the situation in the
geometric theory of defects which is considered in the next sections is
different. In general, only in the final state (after creation of dislocations)
an elastic media occupies the whole Euclidean space $\MR^3$. In the presence
of dislocations, the initial media coordinates $y^i$ usually do not cover the
whole $\MR^3$, because part of media can be removed or, in contrary, added.
Therefore we use Eulerian coordinates related to media points after elastic
deformation and creation of defects.

In the absence of defects, the displacement vector field is assumed to be
a sufficiently smooth vector field in Euclidean space $\MR^3$. The presence of
discontinuities and (or) singularities in the displacement field is interpreted
as existence of defects in an elastic media called {\em dislocations}.

We shall consider only static deformations in what follows. Then basic equations
of elastic media equilibrium for small deformations in Cartesian coordinates
has the form (see., e.g., \cite{LanLif87R}, chapter I, \S 2,4)
\begin{align}                                                     \label{estati}
  \pl_j\s^{ji}+f^i&=0,
\\                                                                 \label{eHook}
  \s^{ij}&=\lm\dl^{ij}\e_k{}^k+2\m\e^{ij},
\end{align}
where $\s^{ji}$ is the stress tensor ($i$-component of the elastic force acting
on the unit area element with the normal $n^j$) which is assumed to be
symmetric.
{\em The tensor of small deformations} $\e_{ij}$ is given by symmetrised
partial derivatives of the displacement vector field:
\begin{equation}                                                  \label{edefte}
  \e_{ij}:=\frac12(\pl_i u_j+\pl_j u_i),
\end{equation}
where lowering and raising of Latin indices is performed using the Euclidean
metric $\dl_{ij}$ and its inverse $\dl^{ij}$. The letters $\lm$ and $\mu$
denote constants characterising elastic properties of media and called
{\em Lame coefficients}. Functions $f^i(x)$ describe total density of
inelastic forces inside media induced, for example, by gravity forces. We assume
in what follows that such forces are absent: $f^i(x)=0$. Equation (\ref{estati})
is the second Newton law for an equilibrium state, and equality (\ref{eHook})
represents the {\em Hooke law}.

Let us look at elastic deformations from the point of view of differential
geometry. From mathematical standpoint, the map (\ref{eeldef}) represents
diffeomorphism of the Euclidean space $\MR^3$ where the Euclidean metric
$\dl_{ij}$ is induced by the pullback of the map $y^i\mapsto x^i$. It means that
in the linear approximation, the deformed metric is
\begin{equation}                                                  \label{emetri}
  g_{ij}(x)=\frac{\pl y^k}{\pl x^i}\frac{\pl y^l}{\pl x^j}
     \dl_{kl}\approx\dl_{ij}-\pl_iu_j-\pl_ju_i=\dl_{ij}-2\e_{ij},
\end{equation}
that is, it is defined by the tensor of small deformations (\ref{edefte}).

In Riemannian geometry, a metric defines the Levi--Civita connection
$\widetilde\Gamma_{ij}{}^k(x)$ (Christoffel's symbols). The corresponding
curvature tensor after an elastic deformation is identically zero
$\widetilde R_{ijk}{}^l(x)=0$, because the curvature of the Euclidean space is
zero and the map $y^i\mapsto x^i$ is a diffeomorphism. The torsion tensor is
also zero by the same reason, since it is set to zero in the observer space.
Thus elastic deformations of media correspond to trivial Riemann--Cartan
geometry because curvature and torsion tensors vanish.
\section{Dislocations                                            \label{sdislo}}
\begin{figure}[htb]
\hfill\includegraphics[width=.6\textwidth]{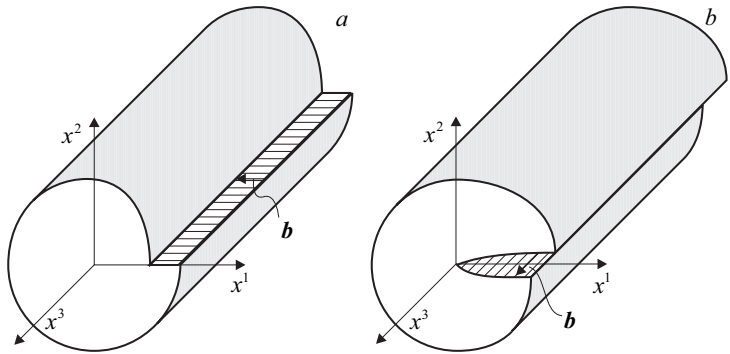}
\hfill {}
\caption{Straight linear dislocations. The edge ({\em a}) and the screw
         ({\em b}) dislocations.}
\label{fdislo}
\end{figure}

We start the description of linear dislocations in elastic media (see., e.g.,
\cite{LanLif87R,Kosevi81R}). The simplest and most usual examples of straight
dislocations are shown in Fig.~\ref{fdislo}. Cut a media along half plane
$x^2=0$, $x^1>0$. Then move the upper part of media above the cut $x^2>0$,
$x^1>0$, on vector $\Bb$ to the axis of dislocation $x^3$ and glue both sides of
the cut. The vector $\Bb$ is called the Burgers vector. In general, the Burgers
vector may be nonconstant on the cut. For the edge dislocation, it varies from
zero to some constant value $\Bb$ as it moves from the dislocation axis. After
the gluing, the media comes to some equilibrium state which is called the edge
dislocation shown in Fig.~\ref{fdislo},\textit{a}. If the Burgers vector is
parallel to the dislocation line, then it is called screw dislocation shown in
Fig.~\ref{fdislo},\textit{b}.

One and the same dislocation can be formed in different ways. For example, if
the Burgers vector is perpendicular to the cut plane and directed from it, the
appeared cavity should be filled with extra media. One can easily imagine that
the appeared defect is the edge dislocation but rotated by the angle $\pi/2$
around $x^3$-axis. This example shows that the dislocation characteristic is not
the cutting surface but the dislocation line or axis (the edge of the cut) and
the Burgers vector.

From topological viewpoint, the media containing several or even infinite number
of dislocations represent the Euclidean space $\MR^3$. In contrast to elastic
deformations, the displacement vector field fails to be a smooth function
because
of cutting surfaces. At the same time we assume that partial derivatives of the
displacement vector $\pl_j u^i$ (the distortion tensor) are smooth functions on
a cutting surface. From physical point of view this assumption is justified
because these derivatives define the deformation tensor (\ref{edefte}). In
its turn, partial derivatives of the deformation tensor should exist and
be continuous functions everywhere in the equilibrium state except, possibly,
the dislocation axis because elastic forces on both sides of the cut must be
equal in the equilibrium state. Since the deformation tensor defines the induced
metric (\ref{emetri}), we assume that the metric and vielbein in $\MR^3$ are
sufficiently smooth functions everywhere except, possibly, dislocation axes.

The main idea of the geometric approach is the following. To describe single
dislocations in the framework of elasticity theory one has to solve equations
for the displacement vector with given boundary conditions. It is possible for
small number of dislocations. But boundary conditions become so complicated for
increasing number of dislocations that this problem becomes unrealistic.
Moreover, one and the same dislocation can be formed by different cuttings which
results in ambiguous displacement vector field. Another disadvantage of this
approach is its inapplicability for description of continuous distribution of
defects because the displacement vector field does not exist in this case since
it has discontinuities at every point. The main variable in the geometric
approach is a vielbein which is a smooth function everywhere except, probably,
dislocation cores. We postulate new equations for vielbein. The transition from
finite number of dislocations to their continuous distribution is natural and
simple in the geometric theory of defects. Singularities in dislocation cores
are smoothed similar to smoothing of masses of point particles after transition
to continuous media.

\begin{figure}[htb]
\hfill\includegraphics[width=.3\textwidth]{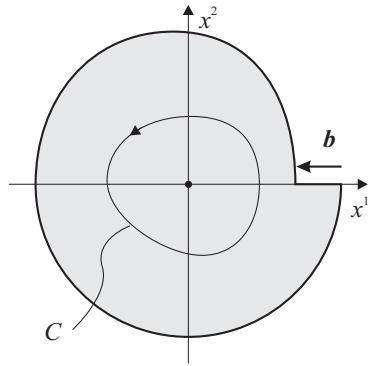}
\hfill {}
 \caption{The section of the media with an edge dislocation. The dislocation
 axis is perpendicular to the figure plane. $C$ is the integration contour for
 the Burgers vector $\Bb$.}
 \label{fburco}
\end{figure}

Now we construct the formalism for the geometric theory of defects. In the
presence of defects, in the equilibrium state, there is no symmetry, and
therefore the notion of distinguished Cartesian coordinates is absent. Hence we
consider an arbitrary curvilinear coordinate system $x^\mu$, $\mu=1,2,3$, in
$\MR^3$. Now Greek letters are used for enumeration of coordinates because we
admit arbitrary coordinate changes. Then the Burgers vector can be expressed by
the integral of the displacement vector
\begin{equation}                                                  \label{eBurge}
  \oint_Cdx^\mu\pl_\mu u^i(x)=-\oint_Cdx^\mu\pl_\mu y^i(x)=-b^i,
\end{equation}
where $C$ is a closed contour surrounding the dislocation axis,
Fig.~\ref{fburco}.
This integral is invariant with respect to arbitrary coordinate changes
$x^\mu\mapsto x^{\mu'}(x)$ and covariant  under $\MS\MO(3)$ global rotations
of $y^i$. Here components of vector field $u^i(x)$ are considered with respect
to orthonormal basis in the tangent space, $u=u^i e_i$.

In the geometric theory of defects, we introduce new independent variable
(vielbein)
\begin{equation}                                        \label{edevid}
  e_\mu{}^i(x):=\begin{cases} ~~~~\pl_\mu y^i, &\text{outside the cut,}\\
               \lim\pl_\mu y^i, &\text{on the cut,}\end{cases}
\end{equation}
instead of partial derivatives of the displacement vector field $\pl_\mu u^i$.
By definition, the vielbein is a smooth function on the cut. Note, that if the
vielbein was simply defined as the partial derivative $\pl_\mu y^i$ then it
would have the $\dl$-function singularity on the cut because functions $y^i(x)$
have a jump there.

The Burgers vector can be represented as the integral over surface $S$ with
contour $C$ as the boundary
\begin{equation}                                                  \label{eBurg2}
  \oint_Cdx^\mu e_\mu{}^i=\int\!\!\int_Sdx^\mu\wedge dx^\nu
  (\pl_\mu e_\nu{}^i-\pl_\nu e_\mu{}^i)=b^i,
\end{equation}
where $dx^\mu\wedge dx^\nu$ is the area element. The definition of vielbein
(\ref{edevid}) implies that the integrand vanish everywhere except the
dislocation axis. The integrand has the $\dl$-function singularity at the
origin for the edge dislocation with constant Burgers vector. The criteria for
the presence of dislocation is the violation of the integrability condition
of the system of equations $\pl_\mu y^i=e_\mu{}^i$:
\begin{equation}                                                  \label{eintco}
  \pl_\mu e_\nu{}^i-\pl_\nu e_\mu{}^i\ne0.
\end{equation}
If dislocations are absent then functions $y^i(x)$ exist and define the
transformation to Cartesian coordinates.

The field $e_\mu{}^i$ is identified with the vielbein in the geometric theory
of defects. Next, compare the integrand in Eq.~(\ref{eBurg2}) with expression
for torsion in Cartan variables
\begin{equation}                                                  \label{ecurcv}
  T_{\mu\nu}{}^i=\pl_\mu e_\nu{}^i-e_\mu{}^j\om_{\nu j}{}^i
                    -(\mu\leftrightarrow\nu).
\end{equation}
They differ only by the terms containing $\MS\MO(3)$-connection
$\om_\mu{}^{ij}=-\om_\mu{}^{ji}$.
This allows us to introduce the following postulate. In the geometric theory of
defects, the Burgers vector corresponding to a surface $S$ is defined by the
integral of the torsion tensor
\begin{equation*}
  b^i:=\int\!\!\int_S dx^\mu\wedge dx^\nu T_{\mu\nu}{}^i.
\end{equation*}
This definition is invariant with respect to general coordinate transformations
of $x^\mu$ and covariant under global rotations. Thus the torsion tensor in the
geometric theory of defects has straightforward physical meaning: it is equal to
the surface density of the Burgers vector.

Physical meaning of $\MS\MO(3)$-connection will be given in Section
\ref{sdiscl}, and now we show how this definition reduces to the expression for
the Burgers vector (\ref{eBurg2}) obtained within the elasticity theory. If the
curvature tensor for the $\MS\MO(3)$-connection vanish, then the connection is
locally trivial and there exist such $\MS\MO(3)$-rotation that
$\om_{\mu i}{}^j=0$. In this case, we return to previous expression
(\ref{eBurg2}).

We have shown that the presence of linear dislocations results in nontrivial
torsion. In the geometric theory of defects, the equality of torsion to zero,
$T_{\mu\nu}{}^i=0$, is naturally considered as the criteria for the absence of
dislocations. Then the term dislocation includes not only linear dislocations
but an arbitrary defects in elastic media. There are also point and surface
defects along with linear dislocations in three dimensions. All of them belong
to dislocations because related to nontrivial torsion.
\section{Disclinations                                           \label{sdiscl}}
Dislocations in elastic media were related to the torsion tensor. To do this, we
introduced the $\MS\MO(3)$-connection. Now we show that the curvature of the
$\MS\MO(3)$-connection defines the surface density of the Frank vector
characterising another well known defects: disclinations in the spin structure
of media \cite{LanLif87R}.

\begin{figure}[htb]
\hfill\includegraphics[width=.6\textwidth]{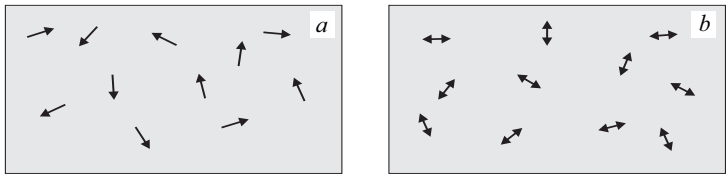}
\hfill {}
\caption{Examples: ferromagnets ({\em a}) and nematic liquid crystals {\em b}).}
\label{fspstr}
\end{figure}

Let a unit vector field $n^i(x)$ $(n^in_i=1)$ be given at every point (spin
structure). For example, $n^i$ has the meaning of magnetic moments at every
point of media for ferromagnets (Fig.~\ref{fspstr}\textit{a}). For nematic
liquid crystals the unit vector field $n^i$ with the equivalence relation
$n^i\sim-n^i$ describes the director field (Fig.~\ref{fspstr}\textit{b}).

Let us fix some direction in media $n_0^i$. Then the field $n^i(x)$ at point
$x$ can be uniquely defined by the angular rotation field
$\om^{ij}(x)=-\om^{ji}(x)$ taking values in the Lie algebra of rotations
$\Gs\Go(3)$ (the rotational angle): $n^i=n_0^j S_j{}^i(\om)$,
where $S_j{}^i\in\MS\MO(3)$ is the rotational matrix corresponding to the
algebra element $\om^{ij}$. We use the following parameterization of the
rotational group by its algebra elements
\begin{equation}                                                  \label{elsogr}
  S_i{}^j=(\ex^{(\om\ve)})_i{}^j=\cos\om\,\dl_i^j
  +\frac{(\om\ve)_i{}^j}\om\sin\om
  +\frac{\om_i\om^j}{\om^2}(1-\cos\om)~          \in\MS\MO(3),
\end{equation}
where $(\om\ve)_i{}^j:=\om^k\ve_{ki}{}^j$ and $\om:=\sqrt{\om^i\om_i}$ is the
modulus of vector $\om^i$. The pseudovector $\om^k=\om_{ij}\ve^{ijk}/2$ where
$\ve^{ijk}$ is the totally antisymmetric third rank tensor, $\ve^{123}=1$,
is directed along rotational axis, its length being equal to the rotation angle.

\begin{figure}[htb]
\hfill\includegraphics[width=.6\textwidth]{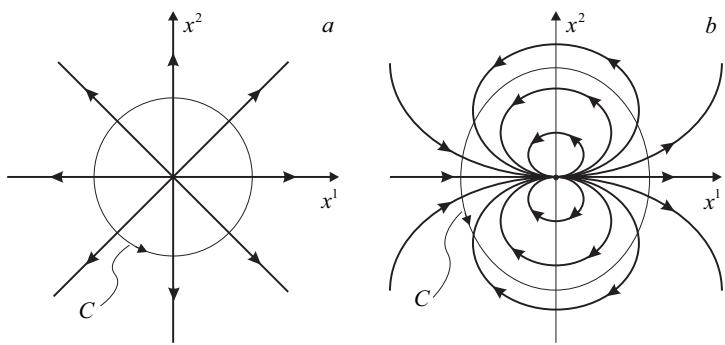}
\hfill {}
\caption{Distribution of unit vector field in the plane $x^1,x^2$ for linear
disclinations parallel to $x^3$ for $\Theta=2\pi$ (\textit{a}) and $\Theta=4\pi$
(\textit{b}).}
\label{fdiscl}
\end{figure}

If media has a spin structure, then it may have defects called disclinations.
For linear disclinations parallel to $x^3$ axis vector field $n$ lies in the
perpendicular plane $x^1,x^2$. The simplest examples of linear disclinations are
shown in Fig.~\ref{fdiscl}.
Each disclination is characterized by the Frank vector
\begin{equation}                                                  \label{etheta}
  \Th_i=\e_{ijk}\Om^{jk},
\end{equation}
where
\begin{equation}                                                  \label{eomega}
  \Om^{ij}=\oint_Cdx^\mu\pl_\mu\om^{ij},
\end{equation}
and the integral is taken along closed contour $C$ around the disclination axis.
The length of the Frank vector is equal to the total rotation angle of field
$n^i$ around the disclination.

Vector field $n^i$ defines the map of the Euclidean space into a sphere
$n:~\MR^3\rightarrow\MS^2$. For linear disclinations parallel to $x^3$ axis
this map is restricted to the map of the plane $\MR^2$ into a circle $\MS^1$.
It is clear that the total rotation angle must a multiple of $2\pi$.

Similar to the case of the displacement vector field for dislocations, the field
$\om^{ij}(x)$ taking values in algebra $\Gs\Go(3)$ is not a continuous function
on $\MR^3$ in the presence of disclinations. Let us make a cut in $\MR^3$
bounded by the disclination axis. Then we can consider the field $\om^{ij}(x)$
as smooth on the whole space without the cut. Assume that all partial
derivatives of $\om^{ij}(x)$ have the same limit on both sides of the cut. Then
define the new field
\begin{equation}                                                  \label{edesoc}
  \om_\mu{}^{ij}:=\begin{cases}~~~~\pl_\mu\om^{ij}, &\text{outside the cut,}\\
                 \lim\pl_\mu\om^{ij}, &\text{on the cut.} \end{cases}
\end{equation}
By construction, functions $\om_\mu{}^{ij}$ are smooth everywhere except,
possibly, disclination axis. Then the Frank vector can be represented by
the surface integral
\begin{equation}                                                  \label{eFrank}
  \Om^{ij}=\oint_Cdx^\mu\om_\mu{}^{ij}=
 \int\!\!\int_Sdx^\mu\wedge dx^\nu(\pl_\mu\om_\nu{}^{ij}-\pl_\nu\om_\mu{}^{ij}),
\end{equation}
where $S$ is an arbitrary surface with boundary $C$. If the field
$\om_\mu{}^{ij}$ is given, then the integrability conditions for the system of
equations $\pl_\mu\om^{ij}=\om_\mu{}^{ij}$ are represented by equalities
\begin{equation}                                                  \label{eomcon}
    \pl_\mu\om_\nu{}^{ij}-\pl_\nu\om_\mu{}^{ij}=0.
\end{equation}
This noncovariant equality yields the criteria of the absence of disclinations.

We identify the field $\om_\mu{}^{ij}$ with the $\MS\MO(3)$-connection in the
geometric theory of defects. The terms with derivatives in the expression for
the curvature
\begin{equation}                                                  \label{ecucav}
  R_{\mu\nu j}{}^i=\pl_\mu \om_{\nu j}{}^i-\om_{\mu j}{}^k
                       \om_{\nu k}{}^i-(\mu\leftrightarrow\nu),
\end{equation}
coincide with these in Eq.~(\ref{eomcon}). Therefore we postulate the covariant
criteria of the absence of disclinations as vanishing of the curvature tensor
for the $\MS\MO(3)$-connection: $R_{\mu\nu}{}^{ij}=0$. At the same time we give
physical meaning of curvature in Cartan variables as the surface density of the
Frank vector:
\begin{equation}                                                  \label{eOmega}
    \Om^{ij}:=\int\!\!\int dx^\mu\wedge dx^\nu R_{\mu\nu}{}^{ij}.
\end{equation}
This definition reduces to the previous expression for the Frank vector
(\ref{eFrank}) in the case when rotations of vector $n$ are confined to a fixed
plane. Then rotations are restricted to the subgroup
$\MS\MO(2)\subset\MS\MO(3)$. The quadratic terms in the expression for the
curvature (\ref{ecucav}) in this case vanish because the rotation group of the
plane $\MS\MO(2)$ is Abelian, and we get the previous expression
(\ref{eFrank}) for the Frank vector.
\section{The 't Hooft--Polyakov monopole}                         \label{smotrp}
The 't Hooft--Polyakov monopole is the static spherically symmetric solution
with finite energy of the field equations of $\MS\MU(2)$-gauge Yang--Mills
model with the triplet of scalar fields $\vf$ in the adjoint representation and
$\lm\vf^4$ interaction \cite{tHooft74,Polyak74R} (see.\ also
\cite{Monast95,Rubako05,Shnir05}). Many other static solutions are related to
this one but does not have spherical symmetry and satisfy some boundary
conditions at infinity where the triplet of scalar fields takes values on
two-dimensional sphere and components of the $\MS\MU(2)$-connection tend to
zero. These solutions are divided into homologically inequivalent classes and
characterised by the topological charge (index of the map $\MS^2\to\MS^2$ of the
boundary of three-dimensional Euclidean space represented by two-dimensional
sphere into the range of values of the scalar fields triplet). These classes of
field configurations have some properties of particles (finiteness of energy,
stability, and localization in space) and are significantly interesting from
theoretical point of view.

It will be shown that solutions of the 't Hooft--Polyakov type have
straightforward physical interpretation in the geometric theory of defects
describing continuous distribution of dislocations and disclinations because
Lie algebras $\Gs\Gu(2)$ and $\Gs\Go(3)$ are isomorphic \cite{Katana20A}.
\subsection{The action and vacuum solutions}
We remind that the Lie algebra $\Gs\Gu(2)$ is compact, simple, coincides
with the Lie algebra of three-dimensional rotations $\Gs\Go(3)$, and defined by
the commutation relations
\begin{equation}                                                  \label{ubbxvh}
  [J_i,J_j]=-\ve_{ij}{}^k J_k,\qquad i,j,k=1,2,3,
\end{equation}
where ${J_i}$ is the Lie algebra basis, $\ve_{ijk}$ is the totally antisymmetric
third rank tensor, and raising and lowering of indices is performed by using
the Euclidean metric $\dl_{ij}$ which is proportional to the Killing--Cartan
form in this case.

Consider the $\MS\MU(2)$-gauge model in Minkowskian space $\MR^{1,3}$ with
Cartesian coordinates $x^\al$ which is described by the following Lagrangian
\begin{equation}                                                  \label{ubcvdt}
  L=-\frac14 F^{\al\bt i}F_{\al\bt i}+\frac12\nb^\al\vf^i\nb_\al\vf_i
  -\frac14\lm\big(\vf^2-a^2\big)^2,
\end{equation}
where $A_\al{}^i$ are components of $\MS\MU(2)$-connection local form
(Yang--Mills fields),
\begin{equation*}
  F_{\al\bt}{}^i:=\pl_\al A_\bt{}^i-\pl_\bt A_\al{}^i
  +eA_\al{}^jA_\bt{}^k\ve_{jk}{}^i
\end{equation*}
is the Yang--Mills field strength (components of the local curvature form of the
$\MS\MU(2)$-connection), $e\in\MR$, $\lm>0$, and $a>0$ are coupling constants,
$\vf:=(\vf^i)\in\MR^3$ is the triplet of real scalar fields transforming under
adjoint representation of $\MS\MU(2)$ group, $\vf^2:=\vf^i\vf_i$, and
$\nb_\al\vf^i:=\pl_\al\vf^i+eA_\al{}^j\vf^k\ve_{jk}{}^i$ is the covariant
derivative of scalar fields.

Since the gauge fields in Eq.~(\ref{ubcvdt}) transform under the adjoint
representation of the $\MS\MU(2)$ group and it coincides with the fundamental
representation of $\MS\MO(3)$ group, everything is reduced to the orthogonal
rotational group $\MS\MO(3)$ from the point of view of equations of motion.

Lagrangian (\ref{ubcvdt}) yields equations of motion:
\begin{align}                                                     \label{ubxbcg}
  \frac{\dl S}{\dl A_\al{}^i}=&\nb_\bt F^{\bt\al}{}_i+e(\nb^\al\vf^j)\vf^k
  \ve_{ikj}=0,
\\                                                                \label{ubbcfd}
  \frac{\dl S}{\dl\vf^i}=&-\nb^\al\nb_\al\vf_i-\lm(\vf^2-a^2)\vf_i=0.
\end{align}

It implies the Hamiltonian density
\begin{equation}                                                  \label{ubbxvf}
  H=-\frac12 P^{\mu i}P_{\mu i}+\frac14F^{\mu\nu i}F_{\mu\nu i}
  +\frac12p^ip_i-\frac12\nb^\mu\vf^i\nb_\mu\vf_i+\frac14\lm\big(\vf^2-a^2\big)^2
  +\mu^i\nb_\mu P^\mu{}_i+\lm^i P^0{}_i,
\end{equation}
where $(P^\al{}_i)=(P^0{}_i,P^\mu{}_i)$ and $p_i$ are momenta conjugate to
potentials $(A_\al{}^i)=(A_0{}^i,A_\mu{}^i)$ and scalar fields $\vf_i$, and
$\mu^i,\lm^i$ are Lagrange multipliers standing in front of constraints of the
first class $P^0{}_i=0$ and $\nb_\mu P^\mu{}_i=0$. We remind that Greek letters
from the middle of the alphabet take only space values $\mu,\nu,\dotsc=1,2,3$.
The energy is, by definition, the numerical value of the Hamiltonian for
physical degrees of freedom, i.e.\ the Hamiltonian after solution of all
constraints and gauge conditions. In the case under consideration, the energy
density for a given fields configuration is obtained from Hamiltonian
(\ref{ubbxvf}) after discarding the last two terms proportional to constraints:
\begin{equation}                                                  \label{uvvxre}
  E=-\frac12 P^{\mu i}P_{\mu i}+\frac14F^{\mu\nu i}F_{\mu\nu i}
 +\frac12p^ip_i-\frac12\nb^\mu\vf^i\nb_\mu\vf_i+\frac14\lm\big(\vf^2-a^2\big)^2.
\end{equation}
It is explicitly positive definite. Remember that the space Greek indices
$\mu,\nu$ are raised and lowered by negative definite metric
$\eta_{\mu\nu}=-\dl_{\mu\nu}$ in our case.

The expression for energy density (\ref{uvvxre}) does not depend on $A_0{}^i$.
For simplicity, we choose the time gauge $A_0{}^i=0$. The solutions of equations
of motion (\ref{ubxbcg}), (\ref{ubbcfd}) with minimal energy correspond to
vacuum. In the considered case, the minimal value equals zero and is achieved if
and only if the following conditions hold:
\begin{equation}                                                  \label{ubnbgt}
  P^\mu{}_i=0,\qquad p_i=0,\qquad F_{\mu\nu}{}^i=0,\qquad\nb_\al\vf^i=0,\qquad
  \vf^2=a^2.
\end{equation}
The first two conditions mean that the vacuum solution in the time gauge must be
static. The third condition imply that components of the gauge fields must be
pure gauge, and, without loss of generality, we put $A_\mu{}^i=0$. Then the last
two equations imply equalities $\pl_\mu\vf^i=0$ and $\vf^2=a^2$.
\subsection{Static spherically symmetric solutions}               \label{snvbfy}
We consider the following ansatz $A_0{}^i=0$, $A_\mu{}^i=A_\mu{}^i(\Bx)$,
$\vf^i=\vf^i(\Bx)$, where $\Bx:=(x^\mu)\in\MR^3$ is a point in Euclidean space.
In this case, equations of motion (\ref{ubxbcg}), (\ref{ubbcfd}) are
\begin{equation}                                                  \label{ubxvcp}
\begin{split}
  \nb_\nu F^{\nu\mu}{}_i+e(\nb^\mu\vf^j)\vf^k\ve_{ikj}=0,
\\
  -\nb^\mu\nb_\mu\vf_i-\lm(\vf^2-a^2)\vf_i=0.
\end{split}
\end{equation}
These are exactly the Euler--Lagrangian equations for the Euclidean
three-dimensional action with the Lagrangian
\begin{equation}                                                  \label{ucxskn}
  L=-\frac14 F^{\mu\nu i}F_{\mu\nu i}+\frac12\nb^\mu\vf^i\nb_\mu\vf_i
  -\frac14\lm\big(\vf^2-a^2\big)^2,
\end{equation}
depending only on the space components $A_\mu{}^i(\Bx)$ and $\vf^i(\Bx)$.

Now we make more precise the definition of spherical symmetry. The rotational
group $\MS\MO(3)$ acts naturally in the coordinate space $(x^\mu)\in\MR^3$, on
which all fields are defined. Moreover, there is the second three-dimensional
Euclidean space $(\vf^i)\in\MR^3$ -- the target space. Therefore the action of
the rotational group should be extended. There is the alternative: we can say
that the $\MS\MO(3)$-group either does not act in the target space at all or it
acts in the same way as in the coordinate space $x^\mu$. The 't Hooft--Polyakov
monopole corresponds to the second definition. In this case, the group of global
$\MS\MO(3)$-rotations acts on Greek and Latin indices in the same way, and they
can be identified.

We change the sign of the space metric $\eta_{\mu\nu}\mapsto\dl_{\mu\nu}$
because Greek and Latin indices are identified in what follows. In other words,
we minimize the energy
\begin{equation}                                                  \label{ubsghj}
  \CE:=\int \!d\Bx\left(\frac14F^{\mu\nu i}F_{\mu\nu i}
  +\frac12\nb^\mu\vf^i\nb_\mu\vf_i+\frac14\lm\big(\vf^2-a^2\big)^2\right),
\end{equation}
where raising and lowering of Greek indices $\mu,\nu=1,2,3$ are performed by the
Euclidean metric $\dl_{\mu\nu}$, and
\begin{equation}                                                  \label{uvvxfd}
\begin{split}
  F_{\mu\nu}{}^i=&\pl_\mu A_\nu{}^i-\pl_\nu A_\mu{}^i
  +eA_\mu{}^jA_\nu{}^k\ve_{jk}{}^i,
\\
  \nb_\mu\vf^i=&\pl_\mu\vf^i+eA_\mu{}^j\vf^k\ve_{jk}{}^i.
\end{split}
\end{equation}
The Euler--Lagrangian equations for functional (\ref{ubsghj}) are
\begin{equation}                                                  \label{uvvxfs}
\begin{split}
  \frac{\dl\CE}{\dl A_\mu{}^i}=&-\nb_\nu F^{\nu\mu}{}_i+e(\nb^\mu\vf^j)\vf^k
  \ve_{ikj}=0,
\\
  \frac{\dl\CE}{\dl\vf^i}=&-\nb^\mu\nb_\mu\vf^i+\lm(\vf^2-a^2)\vf^i=0.
\end{split}
\end{equation}
This system of equations is solved with the spherically symmetric boundary
conditions:
\begin{equation}                                                  \label{ubcdnh}
  \underset{r\to\infty}\lim A_\mu{}^i\to0,\qquad
  \underset{r\to\infty}\lim \vf^i\to \frac{x^i}ra,
\end{equation}

Now we make the spherically symmetric ansatz
\begin{equation}                                                  \label{uncbgf}
  \vf^i=\frac{x^i}r\frac H{er},\qquad
  A_\mu{}^i=\frac{\ve_\mu{}^{ij}x_j}r\frac{K-1}{er},
\end{equation}
where $H(r)$ and $K(r)$ are some unknown functions on radius only. After simple
calculations, the Euler--Lagrange equations (\ref{uvvxfs}) become
\begin{equation}                                                  \label{ubvxgd}
\begin{split}
  r^2K''=&K\big(K^2+H^2-1\big),
\\
  r^2H''=&2HK^2+\lm\left(\frac{H^2}{e^2}-a^2r^2\right)H.
\end{split}
\end{equation}
There is only one analytic solution known at present for $\lm=0$. It is
\cite{PraSom75,Bogomo76}
\begin{equation}                                                  \label{uvxcse}
  K=\frac{ear}{\sh(ear)},\qquad H=\frac{ear}{\tanh(ear)}-1,
\end{equation}
and called the {\em Bogomol'nyi--Prasad--Sommerfield solution}.
It is easily verified that this solution has finite energy.
Numerical analysis of the system of equations (\ref{ubvxgd}) shows that there
exist other spherically symmetric solutions with finite energy.
\subsection{The 't Hooft--Polyakov monopole in the geometric theory of defects}
It was shown in section \ref{snvbfy} that static monopole solutions minimize the
energy (\ref{ubsghj}). This is the three-dimensional functional depending on
$\MS\MO(3)$-connection, in which the metric is supposed to be Euclidean.
Consider it as the expression for the free energy in the geometric theory of
defects, the triplet of scalar fields $\vf^i$ being considered as the source of
defects.

The Euclidean metric means that elastic stresses in media are absent. The
Cartan variables for the monopole solutions are
\begin{equation}                                                  \label{ubbcvl}
  e_\mu{}^i=\dl_\mu^i,\qquad\om_\mu{}^{ij}=A_\mu{}^k\ve_k{}^{ij}
  =(\dl_\mu^jx^i-\dl_\mu^ix^j)\frac{K-1}{er^2},
\end{equation}
where spherically symmetric $\MS\MO(3)$-connection (\ref{uncbgf}) is used.
Note that vielbein is also chosen in the spherically symmetric form. Simple
calculations yield the following expressions for curvature and torsion:
\begin{align}                                                     \label{ubbcvh}
  R_{\mu\nu}{}^k=\frac12R_{\mu\nu}{}^{ij}\ve_{ij}{}^k=F_{\mu\nu}{}^k
  =&\ve_{\mu\nu}{}^k\frac{K'}{er}-\frac{\ve_{\mu\nu}{}^jx_jx^k}{er^3}
  \left(K'-\frac{K^2-1}r\right),
\\                                                                \label{unbgtr}
  T_{\mu\nu}{}^k=&\left(\dl_\mu^kx_\nu-\dl_\nu^kx_\mu\right)\frac{K-1}{er^2}.
\end{align}

In the geometric theory of defects, curvature (\ref{ubbcvh}) and torsion
(\ref{unbgtr}) have physical interpretation as surface densities of Frank and
Burgers vectors. That is they are equal to the $k$-th components of the
corresponding vectors on the unit surface area element $dx^\mu\wedge dx^\nu$.
If $s^\mu$ is the normal to the area element, then the corresponding densities
of Frank and Burgers vectors are:
\begin{align}                                                     \label{uvvxgf}
  f_\mu{}^i:=&\frac12\ve_\mu{}^{\nu\rho}R_{\nu\rho}{}^i
  =\frac1{3er}\dl_\mu^i\left(2K'+\frac{K^2-1}r\right)-\frac1{er}
  \left(\hat x_\mu\hat x^i-\frac13\dl_\mu^i\right)\left(K'-\frac{K^2-1}r\right),
\\                                                                \label{ubfgry}
  b_\mu{}^i:=&\frac12\ve_\mu{}^{\nu\rho}T_{\nu\rho}{}^i
  =\ve_\mu{}^{ij}\hat x_j\frac{K-1}{er},
\end{align}
where $\hat x^\mu:=x^\mu/r$ and the tensor $f_\mu{}^i$ is decomposed into the
irreducible parts.

The functions $K(r)$ and $H(r)$ for the Bogomol'nyi--Prasad--Sommerfield
solution are given by Eqs.~(\ref{uvxcse}). They have the following asymptotics:
\begin{equation}                                                  \label{unnvhf}
\begin{aligned}
  K\big|_{r\to0}\approx & 1-\frac{(ear)^2}6-\frac{(ear)^4}{120}, &
  \qquad K\big|_{r\to\infty}\approx & 2ear\ex^{-ear}\to0,
\\
  H\big|_{r\to0}\approx & 1+\frac{(ear)^2}3-\frac{2(ear)^4}{15},  &
  H\big|_{r\to\infty}\approx & ear-1\to\infty.
\end{aligned}
\end{equation}
The corresponding asymptotics of the densities of the Frank and Burgers vectors
are:
\begin{equation}                                                  \label{uvnfuy}
\begin{split}
  f_\mu{}^i\big|_{r\to0}\approx&-\frac13\dl_\mu^i\left(ea^2+\frac7{90}e^3a^4r^2
  \right)+\frac2{45}x_\mu x^ie^3a^4\to-\frac13\dl_\mu^iea^2,
\\
  b_\mu{}^i\big|_{r\to0}\approx & -\frac16\ve_\mu{}^{ij}x_j\left(ea^2+
  \frac{e^2a^4r^2}{20}\right)\to-\frac16\ve_\mu{}^{ij}x_jea^2,
\\
  \vf^i\big|_{r\to0}\approx & \frac13x^i\left(ea^2-\frac{2e^3a^4r^2}5\right)
  \to\frac13x^iea^2,
\\
  f_\mu{}^i\big|_{r\to\infty}\approx & -\frac{x_\mu x^i}{er^4}\to0,
\\
  b_\mu{}^i\big|_{r\to\infty}\approx & -\ve_\mu{}^{ij}x_j\frac1{er^2}\to0,
\\
  \vf^i\big|_{r\to\infty}\approx & \frac{x^i}r\left(a-\frac1{er}\right)
  \to\frac{x^i}ra.
\end{split}
\end{equation}
It implies, in particular, that the energy integral (\ref{ubsghj}) converges.

Thus monopole solutions of $\MS\MO(3)$-gauge model describe continuous
distribution of dislocations and disclinations in continuous media. The
descriptive picture of this distribution of defects by the displacement vector
field and $n$-field is absent because they are not defined for continuous
distribution of defects.
\section{Spherically symmetric disclinations}                     \label{sbnsgt}
Let as consider the Chern--Simons action for the $\MS\MO(3)$-connection as the
free energy for disclinations \cite{Katana18AR,Katana18BR}. Point disclinations
considered in the present section are described in \cite{KatVol20}.

Consider three-dimensional Euclidean space with Cartesian coordinates
$(x^\mu)\in\MR^3$, $\mu=1,2,3$. Let components of the local
$\MS\MO(3)$-connection form $A_\mu{}^{ij}(x)=-A_\mu{}^{ji}(x)$, $i,j=1,2,3$
(the Yang--Mills fields) be given. From geometric point of view, we may assume
that the Riemann--Cartan geometry is given on topologically trivial
manifold $\MR^3$ defined by the flat vielbein $e_\mu{}^i$ satisfying the
equality $\dl_{\mu\nu}=e_\mu{}^ie_\nu{}^j\dl_{ij}$ and $\MS\MO(3)$-connection
$\om_{\mu i}{}^j=A_{\mu i}{}^j$.

Since there is the third-rank totally antisymmetric tensor $\ve_{ijk}$ in three
dimensions, the connection components can be parameterised by the field with two
indices: $ A_\mu{}^{ij}=A_\mu{}^k\ve_k{}^{ij}$. The related components of the
local curvature form for the $\MS\MO(3)$-connection are
\begin{equation}                                                  \label{ubnhyf}
  F_{\mu\nu k}:=\frac12F_{\mu\nu}{}^{ij}\ve_{ijk}
  =\pl_\mu A_{\nu k}-\pl_\nu A_{\mu k}+A_\mu{}^i A_\nu{}^j\ve_{ijk}.
\end{equation}

We assume that the group of global $\MS\MO(3)$-rotations acts simultaneously
on the base $\MR^3$ and on the Lie algebra $\Gs\Go(3)$, which is also the
three-dimensional space $\MR^3$ as a vector space. It means that if
$S\in\MS\MO(3)$ is an orthogonal matrix, then the transformation has the form
\begin{equation*}
  A_\mu{}^{ij}\mapsto S^{-1\nu}_{~~\mu} A_\nu{}^{kl}S_k{}^iS_l{}^j,\qquad
  S\in\MS\MO(3).
\end{equation*}
The difference between Greek and Latin indices disappears under this assumption,
but we shall distinguish them if possible.

The most general spherically symmetric connection components are
\begin{equation}                                                  \label{unbcht}
  A_\mu{}^i=\ve_\mu{}^{ij}\frac{x_j}r\frac{K-1}r+\dl_\mu^iV(r)
  +\frac{x_\mu x^i}{r^2}U(r),\qquad r\ge0,
\end{equation}
where $K(r)$, $V(r)$, $U(r)$ are arbitrary sufficiently smooth functions on
radius. The case $V=U=0$ corresponds to the 't Hooft--Polyakov monopole
(\ref{uncbgf}).

Straightforward calculations of the components of the spherically symmetric
curvature tensor yield
\begin{multline}                                                  \label{ubcnhy}
  F_{\mu\nu}{}^i=\frac{\ve_{\mu\nu}{}^i}r\big[K'+rV(V+U)\big]
  +\frac{\ve_{\mu\nu}{}^jx_jx^i}{r^3}\left(-K'+\frac{K^2-1}r-rVU\right)+
\\
  +\frac{x_\mu\dl_\nu^i-x_\nu\dl_\mu^i}{r^2}\big[rV'-U-(K-1)(V+U)\big].
\end{multline}

We assume that the free energy expression for the $\MS\MO(3)$-connection is
given by the Chern--Simons action \cite{CheSim74}, which is conveniently written
in differential forms notations,
\begin{equation}                                                  \label{uncbyk}
  S_{\Sc\Ss}:=\int_{\MR^3}\!\!\!\tr\left(dA\wedge A
  -\frac23A\wedge A\wedge A\right)
\end{equation}
where $dx^\mu A_{\mu i}{}^j(x)$ is the matrix-valued connection local $1$-form,
the symbol $\wedge$ denotes external multiplication, and matrix indices are
dropped. The Euler--Lagrange equations for the Chern--Simons action
(\ref{uncbyk}) are nonlinear: $F_{\mu\nu}{}^i=0$ (flat connection). In the
spherically symmetric case, these equations reduce to the following system:
\begin{align}                                                     \label{unbhhu}
  K'+rV(V+U)=&0,
\\                                                                \label{unmvjo}
  -K'+\frac{K^2-1}r-rVU=&0,
\\                                                                \label{ukfoij}
  rV'-U-(K-1)(V+U)=&0,
\end{align}
because the tensor structures in Eq.~(\ref{ubcnhy}) are functionally
independent.
\begin{theorem}
A general solution to the system of equations (\ref{unbhhu})--(\ref{ukfoij}) is
\begin{align}                                                     \label{ubbxvl}
  K=&\cos f,
\\                                                                \label{ujdfki}
  V=&\frac{\sin f}r,
\\                                                                \label{ujgytr}
  U=&\frac{rf'-\sin f}r,
\end{align}
where $f(r)$ is an arbitrary sufficiently smooth function on radius $r\ge0$.
\end{theorem}
The proof is given in \cite{KatVol20}.

Thus a general spherically symmetric solution of the Euler--Lagrange equations
is
\begin{equation}                                                  \label{uncbhf}
  A_\mu{}^i=\frac{\ve_\mu{}^{ij}x_j}{r^2}(\cos f-1)+\dl_\mu^i\frac{\sin f}r
  +\frac{x_\mu x^i}{r^3}(rf'-\sin f),
\end{equation}
where $f(r)$ is an arbitrary function. If function $f$ is smooth and tends to
zero sufficiently fast as $r\to0$, then the curvature for the
$\MS\MO(3)$-connection is identically zero on the whole $\MR^3$, and there is no
disclination. If $f(0)\ne0$, then disclinations may appear at the origin of
the coordinate system. To understand their structure, we have to find the unit
vector field $n(x)$.
\subsection{Point disclinations}
In the simply connected domains of the Euclidean space with vanishing curvature,
the connection components are pure gauge $A_\mu=\pl_\mu S^{-1}S$, where
$S\in\MS\MO(3)$ and matrix indices are dropped. Our aim is to find the
orthogonal matrix $S$ for a given connection (\ref{uncbhf}). The equations for
$S$ has the form $\pl_\mu S^{-1}=A_\mu S^{-1}$ and coincides with the condition
of parallel displacement of vectors. The parallel displacement does not depend
on the curve along which it is moved. Therefor we consider the curve
$\g=x(t)$, $t\in[0,b]$ with the beginning at point $x_0:=x(0)$ and the end at
point $x_b:=x(b)$. Then we obtain the ordinary differential equation for matrix
$S$ along $\g$:
\begin{equation}                                                  \label{ujfkiy}
  \dot S^{-1}=\dot x^\mu A_\mu S^{-1}.
\end{equation}
When the curve goes through the point $x(t)$, the solution of this equation
is given by the path ordered exponent:
\begin{equation}                                                  \label{unbhcy}
  S^{-1}\big(x(t)\big)=\P\exp\left(\int_0^t\!\!\!ds\,\dot
  x^\mu(s)A_\mu(s)\right)S^{-1}_0,
\end{equation}
where $S_0$ is the orthogonal matrix at the initial point $x_0$.

Let $\g$ be the ray starting at the infinite point and ending at point $x$,
i.e.\ $\g=(x^\mu t)$, $t\in[1,\infty]$, and $S_0:=S(\infty):=\one$. Then the
equality $\dot x^\mu A_{\mu i}{}^j=f'x^k\ve_{ki}{}^j$ holds for connection
(\ref{uncbhf}). Now we can easily check that exponents under the integral
commute: $[\dot x^\mu A_\mu,\dot x^\nu A_\nu]=0$. Consequently, the path
ordered exponent coincides with the usual one, and the integral (\ref{unbhcy})
can be easily taken:
\begin{equation*}
  \int_\infty^1\!\!\!ds\,\dot x^\mu A_{\mu i}{}^j
  =\int_\infty^1\!\!\!ds\,f'x^k\ve_{ki}{}^j
  =\int_\infty^1\!\!\!ds\,\frac{df}{d(rs)}x^k\ve_{ki}{}^j
  =\frac{x^k\ve_{ki}{}^j}r\big[f(r)-f(\infty)\big].
\end{equation*}
That is the solution of Eq.~(\ref{ujfkiy}) is
\begin{equation}                                                  \label{unnvmf}
  S^{-1j}_{~~i}=\exp(-f^k\ve_{ki}{}^j),\qquad
  f^k:=\frac{x^k}r\big[f(\infty)-f(r)\big].
\end{equation}
Vector $(f^k)$ is an element of Lie algebra $\Gs\Go(3)$. Its direction coincides
with the rotational axis in the isotopic space and its length is equal to the
rotation angle. The exponential map for the $\MS\MO(3)$-group is well known:
\begin{equation}                                                  \label{unncbg}
  S_i{}^j=\exp(f^k\ve_{ki}{}^j)
  =\dl_i^j\cos F+\frac{f^k\ve_{ki}{}^j}F\sin F
  +\frac{f_if^j}{F^2}(1-\cos F),
\end{equation}
where $F^2:=f^if_i=\big[f(\infty)-f(r)\big]^2$.
\subsection{Examples of point disclinations}
The rotational matrix (\ref{unncbg}) is defined by the difference
$f(\infty)-f(r)$, where $f(r)$ is an arbitrary sufficiently smooth function.
Without lass of generality we put $f(\infty)=0$ and change the sign of $f(r)$.
Then we can choose $F(r)=f(r)$, and the spherically symmetric rotational matrix
takes the form
\begin{equation}                                                  \label{unnpbg}
  S_i{}^j=\exp(f^k\ve_{ki}{}^j)
  =\dl_i^j\cos f+\frac{f^k\ve_{ki}{}^j}f\sin f
  +\frac{f_if^j}{f^2}(1-\cos f),
\end{equation}
where $f^i=x^i f(r)/r$ with arbitrary function $f(r)$, which is equal to zero at
infinity.

\begin{figure}[hbt]
\hfill\includegraphics[width=.3\textwidth]{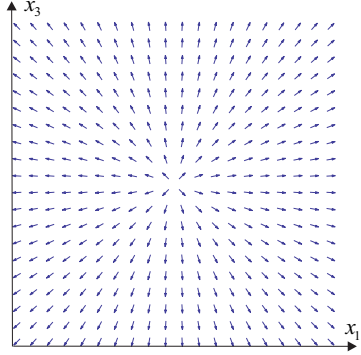}
\hfill {}
\centering\caption{Spherically symmetric ``hedgehog'' disclination. The section
$x_2=0$ is shown in the figure.}
\label{fdisclhedgehog}
\end{figure}

\begin{exa} [\bf ``Hedgehog'' disclination]
Let us choose the spherically symmetric boundary condition at infinity:
$n^i(r=\infty)=x^i/r$. Then the $n$-field has the same form in the whole space
$\MR^3$: $n^i(r):=n^j(\infty)S_j{}^i(r)=x^i/r$ for arbitrary function $f$.
We see that $n$-field is directed along the radius everywhere and has the unit
length. The distribution of the $n$-field is shown in Fig.~\ref{fdisclhedgehog}.
\qed\end{exa}

Now we consider spherically nonsymmetric disclinations. Fix the vector
$n_0:=(0,0,1)$ at infinity, and thus break the spherical symmetry. Then
components of the $n$-field are
\begin{equation*}
\begin{split}
  n_1=&-\frac{x_2}r\sin f+\frac{x_1x_3}{r^2}(1-\cos f),
\\
  n_2=&~~\frac{x_1}r\sin f+\frac{x_2x_3}{r^2}(1-\cos f),\qquad
  n_3=\cos f+\frac{x_3^2}{r^2}(1-\cos f),
\end{split}
\end{equation*}
where coordinate indices are lowered for simplicity to distinguish them from
exponents. Afterwards we go to the spherical coordinates,
$(x_1,x_2,x_3)\mapsto(r,\theta,\vf)$. Then components of the $n$-field are
\begin{equation*}
\begin{split}
  n_1=&-\sin\theta\sin\vf\sin f+\sin\theta\cos\theta\cos\vf(1-\cos f),
\\
  n_2=&~~\sin\theta\cos\vf\sin f+\sin\theta\cos\theta\sin\vf(1-\cos f),\qquad
  n_3=\cos f+\cos^2\theta(1-\cos f).
\end{split}
\end{equation*}
It implies that the limit of $n$-field at the origin does not depend on the path
along which the limit $r\to0$ is taken if and only if when $f(0)=0,\pi$. This is
the degenerate case, when the $n$-field is continuous at zero, and disclinations
are absent. If $f(0)\ne0,\pi$, then the limit $n$-field does depend on the path
along which it goes to the origin. Consequently, in a general case, the origin
of the coordinate system is an essential singularity, and the model describes
point disclinations located at the origin.

After fixing the $n_0$ vector, the invariance under rotations in the $x_1,x_2$
plane remains. Therefore it is sufficient to put $x_2=0$ for visualisation of
disclinations, that is to analyse the distribution of the $n$-field in the
$x_1,x_3$ plane:
\begin{equation*}
  n_1=\frac{x_1x_3}{r^2}(1-\cos f),\qquad
  n_2=\frac{x_1}r\sin f,\qquad
  n_3=\cos f+\frac{x_3^2}{r^2}(1-\cos f).
\end{equation*}
We see that the $n$ vector has nonzero component in the direction perpendicular
to the $x_1,x_3$ plane in general, which slightly obscures the pictures.

Various distributions of the $n$-field depend on the choice of function $f(r)$.
We put $f(\infty)=0$. Then the $n$-field coincides with $n_0$ at infinity. If
$f(0)=0,\pi$, the unit vector field is continuous at zero, and disclinations are
absent. In the opposite case, there are disclinations with an essential
singularity at the origin.
\begin{figure}[hbt]
\hfill\includegraphics[width=.7\textwidth]{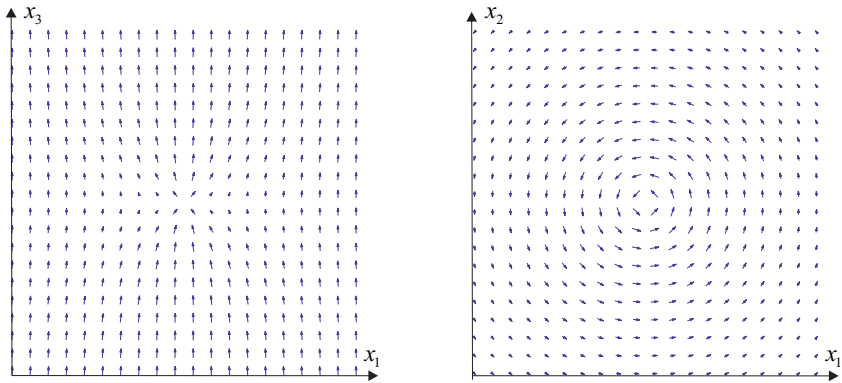}
\hfill {}
\centering\caption{Two sections ($x_2=0$ and $x_3=0$) of the disclination for
$f(r):=\pi\ex^{-r}/2$. The arrows are projections of $n$ vector on corresponding
plane. If the length of an arrow is less then unity, then the vector has the
component in the perpendicular direction. The spherical symmetry is broken by
the boundary condition $n(\infty):=(0,0,1)$.}
\label{fdisclxzxy}
\end{figure}
\begin{exa}
Set $f(r):=\frac\pi2\ex^{-r}$, which implies $f(0)=\frac\pi2$  and
$f(\infty)=0$.
In this case, the vector field $n$ in the plane $x_2=0$ has all three
nontrivial components. Therefore we draw the projections of $n$-field on two
planes  $x_2=0$ and $x_3=0$ in Fig.~\ref{fdisclxzxy} for visualisation. The
projection of the vector field has the unit length on the plane $x_2=0$ at
infinity, because the perpendicular component is absent. The projection
becomes less in internal points because the perpendicular component arises.
The projections of vectors $n$ on the plane $x_3=0$, conversely, are zero at
infinity and nontrivial at internal points which is clear from the picture.
\qed\end{exa}

The equilibrium equations for described disclinations hold everywhere in
$\MR^3$ except the origin of the coordinate system where $\MS\MO(3)$-connection
is singular. The analysis of this singularity is difficult in general because
equations are nonlinear, and we postpone it for further investigations. We
consider linear disclinations in the next section, for which equations become
linear. The singularity for these disclinations is proportional to the
$\dl$-function with the support located along the line of disclination.
\section{Linear disclinations}                                    \label{sbcvfg}
We assume that the expression for the free energy is given by the Chern--Simons
action (\ref{uncbyk}) as in the previous section. To describe linear
disclinations  we add the source term to the action
\begin{equation}                                                  \label{qhwgtr}
  S_{\Sc\Ss}[A]+S_\text{int}=\int_{\MR^3}\left(\frac12dA^i\wedge A_i
  +\frac16 A^i\wedge A^j\wedge A^k\epsilon_{ijk}-A^i\wedge J_i \right),
\end{equation}
where $J$ is the 2-form of the disclinations source which is not specified here.
The interaction term is similar to the minimal coupling of the electric charge
to the electromagnetic field in electrodynamics. The equilibrium equations for
action (\ref{qhwgtr}) are
\begin{equation}                                                  \label{uvbxvr}
  F_{\mu\nu}{}^k=J_{\mu\nu}{}^k,
\end{equation}
where $J_{\mu\nu}{}^k$ are components of the source for the
$\MS\MO(3)$-connection.

The first two terms in the action (\ref{qhwgtr}) change by the external
differential under local $\MS\MO(3)$-rotations. Therefore we have to impose the
condition $DJ^k=0$, where $DJ^k:=dJ^k+J^j\wedge\om_j{}^k$ is the external
covariant derivative, for the self-consistency of the Euler--Lagrange equations.

Consider one linear disclination $q^\mu(t)\in\MR^3$, where $t\in\MR$ is a
parameter along the core of disclination. The interaction term is written in the
form
\begin{equation}                                                  \label{ujdght}
  S_\text{int}:=\int dq^\mu A_{\mu i}J^i=\int dt\,\dot q^\mu A_{\mu i} J^i.
\end{equation}
This action is invariant with respect coordinate changes in $\MR^3$ (up to
boundary terms) and arbitrary reparameterization of the curve $q^\mu(t)$.
We assume that the disclination is located in such a way that inequality
$\dot q^3\ne0$ holds everywhere. The three dimensional $\dl$-function is
inserted into the integrand for variation of this action with respect to the
$\MS\MO(3)$-connection:
\begin{equation*}
  S_\text{int}=\int dt d^3x \dot q^\mu A_{\mu i}J^i\dl^3(x-q)=
  \int d^3x\frac{\dot q^\mu}{\dot q^3}A_{\mu i}J^i\dl^2(x-q),
\end{equation*}
where we integrated over $t$ using one $\dl$-function
$\dl\big(x^3-q^3(t)\big)$ and $\dl^2(x-q):=\dl(x^1-q^1)\dl(x^2-q^2)$ denotes
the two-dimensional $\dl$-function on the plane $x^1,x^2$. Then variation of the
interaction term is
\begin{equation}                                                  \label{ubcvdf}
  \frac{\dl S_\text{int}}{\dl A_{\mu i}}=\frac{\dot q^\mu}{\dot q^3} J^i
  \dl^2(x-q).
\end{equation}

We consider equation (\ref{uvbxvr}) on topologically trivial manifold
$\MM\approx\MR^3$ with Cartesian coordinate system $x^1=x$, $x^2=y$ and $x^3$.
The disclination is supposed to be straight and coinciding with the $x^3$ axis,
i.e.\ $q^1=q^2=0$ and $q^3=t$. We are searching solutions of
Eqs.~(\ref{uvbxvr}) which are invariant with respect to translations along $x^3$
axis and describe rotations only in the $x,y$ plane. In this case, the
$\MS\MO(3)$-connection has only two nontrivial components $A_x{}^3$ and
$A_y{}^3$ depending on a point on the $(x,y)\in\MR^2=\MC$ plane. To find the
solution we introduce the complex coordinate $z:=x+iy$. Then two real
components of the $\MS\MO(3)$-connection are united into the complex one:
\begin{equation}                                                  \label{unvbgf}
  A_z{}^3=\frac12A_x{}^3-\frac i2A_y{}^3,\qquad
  A_{\bar z}{}^3=\frac12A_x{}^3+\frac i2A_y{}^3.
\end{equation}
The corresponding curvature tensor (field strength) has only one linearly
independent complex component
\begin{equation}                                                  \label{ubcvdg}
  F_{z\bar z}{}^3=2(\pl_zA_{\bar z}{}^3-\pl_{\bar z}A_z{}^3),
\end{equation}
which is linear in connection. It is the consequence of the fact that the
rotational $\MS\MO(2)$ group acting on the $x,y$ plane is Abelian and nonlinear
terms in the curvature tensor disappear.

In our case, the quadratic terms in the curvature identically disappear, and we
are able to consider sources of the $\dl$-function form because equilibrium
equations (\ref{uvbxvr}) become linear. Now we fix the sources
\begin{equation}                                                  \label{ubdyui}
  F_{z\bar z}{}^3=4\pi i D\dl(z),\qquad D\in\MR,
\end{equation}
where $\dl(z)$ is the two-dimensional $\dl$-function on the complex plane. It is
clear that this source has rotational symmetry.

The solution of equation (\ref{ubdyui}) describes new type of geometric
singularity. If this equation was considered as the second order equation for
the metric then its solution would describe conical singularity on the $x,y$
plane. In this case, the solution describes the wedge dislocation in the
geometric theory of defects \cite{KatVol92}. The situation is now different. We
consider this equation as the first order one for the $\MS\MO(3)$-connection
and show that it describes the defect of the unit vector field (disclination),
the metric being Euclidean.

Equation (\ref{ubdyui}) has the solution
\begin{equation}                                                  \label{uxbvpo}
  A_z{}^3=-\frac{iD}z,\qquad A_{\bar z}{}^3=\frac{iD}{\bar z}.
\end{equation}
To check that this is indeed the solution, one can use the well known formula
(see, e.g.\ \cite{Vladim71}):
\begin{equation}                                                  \label{ubvcfr}
  \pl_z\frac1{\bar z}=\pi\dl(z)\qquad\Leftrightarrow\qquad \pl_{\bar z}\frac1z
  =\pi\dl(z).
\end{equation}

The corresponding real components are
\begin{equation}                                                  \label{unbcyg}
  A_x{}^3=-\frac{2Dy}{x^2+y^2},\qquad A_y{}^3=\frac{2Dx}{x^2+y^2}.
\end{equation}

Outside the $x^3$ axis the curvature is flat, and therefore the connection is
given by partial derivatives of some function. It is the angular rotation field
$\theta(x,y)$ of the unit vector field on the plane in the geometric theory of
defects. This field must satisfy the following system of equations:
\begin{equation}                                                  \label{uxncrt}
  \pl_x\theta=-\frac{2Dy}{x^2+y^2},\qquad \pl_y\theta=\frac{2Dx}{x^2+y^2}.
\end{equation}
The integrability conditions of this system $\pl_{xy}\theta=\pl_{yx}\theta$ are
fulfilled, and one can easily write down a general solution
\begin{equation}                                                  \label{usgwed}
  \theta=-2D\arctan \frac xy+C,\qquad C=\const.
\end{equation}
Fix the constant of integration $C:=\pi D$. Then the solution takes the form
\begin{equation}                                                  \label{unbvtr}
  \tan \frac \theta{2D}=\frac yx=\tan\vf,
\end{equation}
where $\vf$ is the usual polar angle on the $(x,y)\in\MR^2$ plane. If one goes
around $x^3$ axis along contour $C$, then the polar angle changes by $2\pi$.
We have to impose the quantization condition $D=n/2$, $n\in\MZ$ in order that
the rotation angle field $\theta(x,y)$ be well defined.

Thus the angular rotational field takes the form $\theta=n\vf$, where $\vf$ is
the usual polar angle on the $x,y$ plane. It is defined everywhere except the
cut on the half plane, say, $y=0$, $x\ge 0$. The corresponding
$\MS\MO(3)$-connection has only two nontrivial components:
\begin{equation*}
  A_x{}^{12}=-\frac{ny}{x^2+y^2}=-\frac nr\sin\vf,\qquad
  A_y{}^{12}=\frac{nx}{x^2+y^2}=~~\frac nr\cos\vf,
\end{equation*}
where $r:=\sqrt{x^2+y^2}$ is the polar radius. It is defined everywhere on the
$x,y$ plane except the origin where its rotor has $\dl$-function singularity
(\ref{ubdyui}). We see that the $\MS\MO(3)$-connection has much better behaviour
then the respective angular rotation field as it should be in the geometric
theory of defects.

Thus, after going around the $x^3$ axis along the closed contour $C$ the
rotation angle field changes from 0 to $2\pi n$ where $|2\pi n|=|\Om|$ is the
modulus of the Frank vector. It is exactly the linear disclination of the unit
vector field with the core coinciding with the $x^3$ axis. For $n=0$ the
disclination is absent. This case requires separate treatment: for $D=0$, the
equality $\theta=0$ must hold as the consequence of Eq.~(\ref{usgwed}). Two
simplest examples of linear disclinations for $n=1$ and $n=2$ are shown in
Fig.~\ref{fdiscl}, where the distribution of the angular rotation fields are
shown on the $x,y$ plane.
\section{Conclusion}
We give a review of the geometric theory of defects in this paper. Examples of
the known until now disclinations are described. Since the Lie algebras
$\Gs\Go(3)$ and $\Gs\Gu(2)$ are isomorphic, static solutions of $\MS\MU(2)$
gauge models have straightforward physical interpretation in the framework of
the geometric theory of defects. In particular, the 't Hooft--Polyakov monopole
has physical interpretation in crystals describing continuous distributions
of dislocations and disclinations.

We showed that the Chern--Simons action is well suited for description of
single disclinations in the geometric theory of defects. The most general
spherically symmetric $\MS\MO(3)$-connection containing one arbitrary function
of radius is found for description of point disclinations. Two examples are
given: spherically symmetric disclination of the hedgehog form and the point
disclination with constant value of the $n$-field at infinity and essential
singularity ar the origin. The Chern--Simons action describes also linear
disclinations. As an example, we consider straight linear disclinations with the
Frank vector which is a multiple of $2\pi$.

Acknowledgments. The reported study was funded by RFBR, project number
19-11-50067.

\end{document}